\documentclass[conference]{IEEEtran}
\IEEEoverridecommandlockouts

\usepackage{cite}
\usepackage{amsmath,amssymb,amsfonts}
\usepackage{algorithmic}
\usepackage{graphicx}
\usepackage{makecell}
\usepackage{textcomp}
\usepackage{url}
\usepackage{xcolor}
\def\BibTeX{{\rm B\kern-.05em{\sc i\kern-.025em b}\kern-.08em
    T\kern-.1667em\lower.7ex\hbox{E}\kern-.125emX}}
\begin{document}

\title{Streaming Endpointer for Spoken Dialogue using Neural Audio Codecs and Label-Delayed Training\\



\author{
Sathvik Udupa\textsuperscript{1},
Shinji Watanabe\textsuperscript{2},
Petr Schwarz\textsuperscript{1},
Jan Cernocky\textsuperscript{1} \\
\textsuperscript{1}\textit{Brno University of Technology, Czechia}\\
\textsuperscript{2}\textit{Carnegie Mellon University, United States} \\
\{udupa, schwarzp, cernocky\}@fit.vut.cz, shinjiw@ieee.org
}

}

\maketitle



\begin{abstract}
Accurate, low-latency endpointing is crucial for effective spoken dialogue systems. While traditional endpointers often rely on spectrum-based audio features, this work proposes real-time speech endpointing for multi-turn dialogues using streaming, low-bitrate Neural Audio Codec (NAC) features, building upon recent advancements in neural audio codecs. To further reduce cutoff errors, we introduce a novel label delay training scheme. At a fixed median latency of 160 ms, our combined NAC and label delay approach achieves significant relative cutoff error reductions: 42.7\% for a single-stream endpointer and 37.5\% for a two-stream configuration, compared to baseline methods. Finally, we demonstrate efficient integration with a codec-based pretrained speech large language model, improving its median response time by 1200 ms and reducing its cutoff error by 35\%.
\end{abstract}


\section{Introduction}
Advancements in spoken dialogue systems \cite{ji2024wavchat} are leading to widespread adoption of speech technologies.  Consequently, speech is a primary way users interact with applications ranging from voice assistants \cite{faruk2024review} and speech Large Language Models (LLM) \cite{team2023gemini, gpt4o} to spoken dialogue systems like customer support and emergency services. For such applications, it is beneficial to know when a user has stopped speaking so that further downstream processing can begin. 

This process is referred to as speech \textit{endpointing} \cite{Shannon_Simko_Chang_Parada_2017}. While endpointing is related to the broader field of turn-taking prediction \cite{ekstedt22_interspeech, talking_turns}, these tasks address different challenges. Turn-taking encompasses various conversational phenomena such as backchannel prediction, and interruption handling, typically evaluated at fixed latency constraints. In contrast, endpointing focuses specifically on detecting the completion of a user's utterance and optimises for the critical trade-off between response latency and detection accuracy across varying temporal thresholds.
Developing an effective endpointer requires balancing this latency-accuracy trade-off. Erroneous endpoint predictions can result in incomplete data being passed to downstream processes, leading to a degraded user experience \cite{ward05_interspeech}.


Voice Activity Detection (VAD) \cite{sohn1999statistical, 8278160, chang2018temporal} deals with predicting each audio frame as speech or silence. VAD with a fixed delay \cite{hariharan2001robust} represents a conventional approach to endpointing. However, this methodology leads to unnecessary latency and bad user experience, particularly in multi-turn dialogue scenarios. Thus, endpointers with minimal latency \cite{Shannon_Simko_Chang_Parada_2017} are trained with frame-level targets for speech and end-of-speech. The endpoint is detected when the end-of-speech probability exceeds a predefined threshold. In this work, we build a streaming endpointer for multi-turn conversations.


Traditionally, endpointing systems have often processed a single audio stream for the user \cite{Shannon_Simko_Chang_Parada_2017, maas2018combining, liang2023dynamic}. While simpler, this approach can face challenges in complex multi-turn dialogues where distinguishing between user and system speech is important. More recent advancements in turn-taking prediction \cite
{ekstedt22_interspeech} leverage multi-stream architectures, processing user and system audio through separate channels to better model these interactive dynamics. Given these differing approaches and the demands of multi-turn dialogue, a key aspect of our work involves evaluating our proposed methods across both single-stream and two-stream endpointing models.

We propose to use neural audio codecs (NAC) \cite{soundstream, encodec, audiodec, moshi} as speech feature representation for the endpointing task. Neural audio codecs, originally developed for audio compression, provide discrete audio representations. These typically incorporate a quantization-based bottleneck, enabling audio to be represented as discrete codes. These features are nowadays used for a wide range of speech tasks \cite{wu2024codecsuperbindepthanalysissound, shi2024espnetcodeccomprehensivetrainingevaluation} including Automatic Speech Recogniser (ASR) \cite{codec_asr}, speech LLM \cite{moshi} and speech translation \cite{labiausse2025highfidelitysimultaneousspeechtospeechtranslation}. Compared to traditional speech features such as Mel-Spectrograms, neural audio codecs can represent audio at low bit-rates \cite{soundstream} while maintaining high reconstruction quality. Unlike non-streaming, higher bit-rate self-supervised learning (SSL) features \cite{mohamed2022self}, streaming NACs \cite{soundstream, encodec, audiodec, moshi, ahn2024hilcodec} are uniquely suited for real-time applications. Furthermore, utilising neural codec-based endpointer offers multi-task capability, enabling integration of codec-based endpointer with codec-based ASRs and codec-based speech LLMs.


    While using neural codecs for endpointing has benefits, using it can lead to a sub-optimal performance \cite{maas2018combining} mainly due to the presence of short pauses and filler words \cite{hwang2020end}. Utilization of speech features only may prove insufficient to prevent the premature triggering of the endpointer due to silences in the middle of a turn (mid-silence). Several approaches have been explored to enhance endpointer accuracy, including the use of ASR features \cite{maas2018combining}, multi-task training with ASR \cite{chang2019joint, chang22_interspeech, li2020towards, sudo2022streaming, raju2023two, eos2023, bijwadia23_interspeech}, voice activity projection \cite{ekstedt22_interspeech, inoue-etal-2024-multilingual, inoue-etal-2025-yeah} and SSL features along with LLMs \cite{ssl_llm_2024}. Additionally, many works have used additional pause/silence labels \cite{chang2019joint, chang22_interspeech, yoshimura2020end, bijwadia23_interspeech} and pause modelling-based regression \cite{liang2023dynamic}. 

\begin{figure*}[t]
    \centering
    \includegraphics[width=\linewidth]{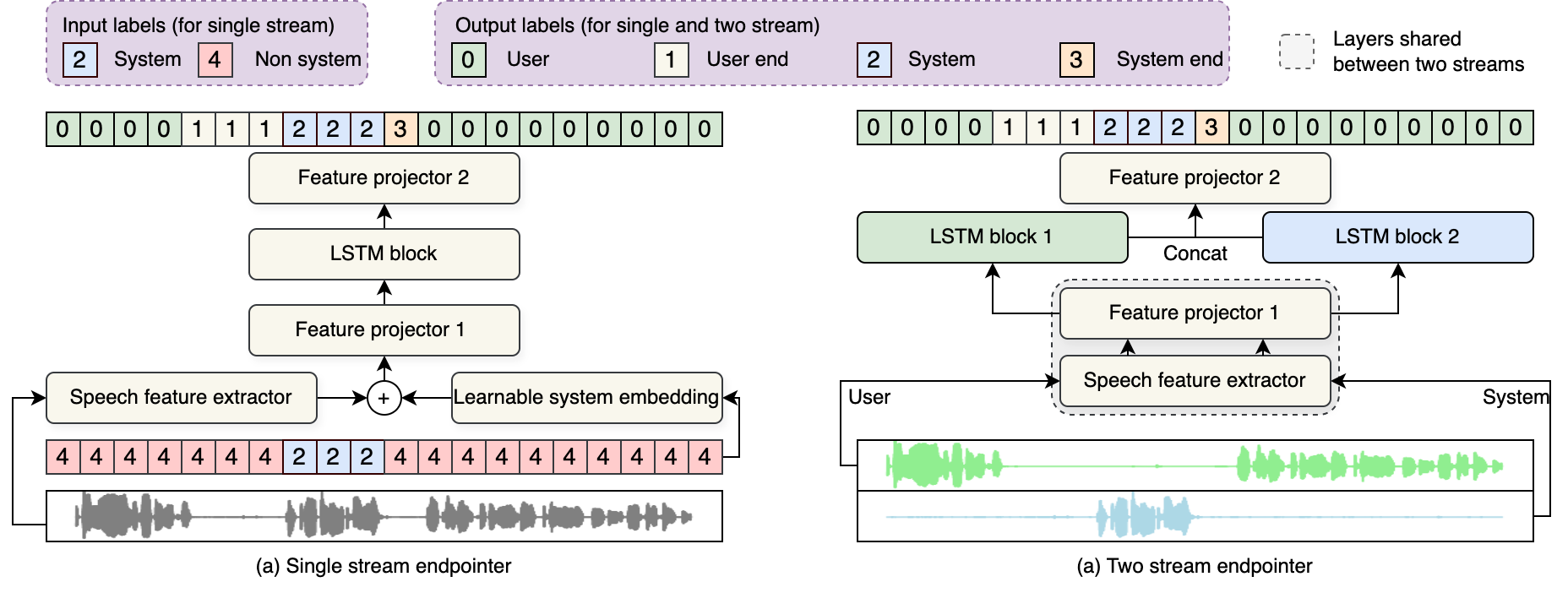}
    \vspace{-8mm}
    \caption{Endpointing model architectures. (a) Single-stream endpointer: Processes a single, merged audio input. It incorporates system activity information (input labels: '2' for System, '4' for Non-system) via a learnable system embedding. (b) Two-stream endpointer: Processes user and system audio through separate streams, utilizing shared initial feature extraction layers (Speech feature extractor and Feature projector 1). Both models perform frame-level 4-class classification (Output labels: 0-User, 1-User end, 2-System, 3-System end). The baseline model uses Mel-spectrograms as the 'Speech feature extractor,' while models using Neural Audio Codecs (NAC) use the NAC encoder.}
    \label{fig:data-labels}
    \vspace{-4mm}
\end{figure*}


    We propose label delay training, a novel approach to reduce errors in standalone speech endpointers without additional features or pause labels. Recent multi-stream speech language models have used delayed tokens \cite{kharitonov-etal-2022-text,copet2024simple,moshi} to reduce dependencies between parallel output streams. In contrast, we apply label delay to a single output stream, specifically to optimize the latency-accuracy trade-off and enhance endpointer performance.
    By shifting target labels, we encourage delayed, high-confidence predictions, which leads to inherently fewer errors. 
    


\noindent
The contributions of our work are as follows - 

\begin{itemize}
    \item Propose the use of streaming neural audio codec (NAC) features for real-time endpointing in multi-turn spoken dialogue.
    \item Introduce a novel label delay training scheme that improves endpointing accuracy by managing the latency-error trade-off.
    \item Provide a comprehensive comparison of single-stream and two-stream endpointing architectures.
    \item Demonstrate integration of a codec-based endpointer with a pretrained codec-based speech LLM, significantly reducing response latency and cutoff errors.
    \item Release data preparation, training, and evaluation code for reproducibility.\footnote{Link omitted for blind submission; will be updated post-review.}
\end{itemize}




\section{Proposed methodology}
In this section, we will first present an overview of our single-stream and two-stream baseline endpointer configurations. Following it, we will elaborate on our proposed approaches - application of neural audio codec features and label delay training for endpointing.

\subsection{Baseline endpointer}
\label{sec:base_ep}

For streaming endpointing, we use Long Short-Term Memory (LSTM) model, trained for frame-level predictions of 4 labels - \texttt{$<$user$>$}, \texttt{$<$user-end$>$}, \texttt{$<$system$>$}, \texttt{$<$system-end$>$}. Endpoint triggering is based on a threshold applied to the turn-end probability. Our baseline utilizes a $40$-channel Mel-Spectrogram feature as a feature extractor at a $25$ Hz frame rate.
We run our experiments on two input configurations - 

\textbf{Single-stream endpointer:} 
\begin{itemize}
\item Uses a single LSTM block on merged user and system audio channels as shown in Figure 1(a).
\item To distinguish between speakers in the merged stream, it utilises explicit system activity labels ('system' vs. 'non-system') as an input feature during both training and inference.\footnote{In practice, these labels can be derived from application-specific cues such as system microphone status or special tokens from a speech LLM.}
\item A learnable embedding corresponding to these system activity labels is added to the acoustic input features, enabling the LSTM to better differentiate between user and system frames.
\end{itemize}

\textbf{Two-stream endpointer:} 
\begin{itemize}
    \item The user and system audio is available in separate audio channels.
    \item The two streams initially pass through shared feature extraction and projection layers. Subsequently, each stream is processed by a dedicated LSTM block to model its specific temporal dynamics as shown in Figure \ref{fig:data-labels}(b).
    \item This architecture does not require explicit system activity labels as input. The separate processing paths and dedicated LSTMs inherently allow the model to distinguish between user and system activity.
\end{itemize}

\subsection{Neural audio codecs}
\label{sec:codecs}
\begin{figure}[b]
    \centering
    \vspace{-4mm}
    \includegraphics[width=1\linewidth]{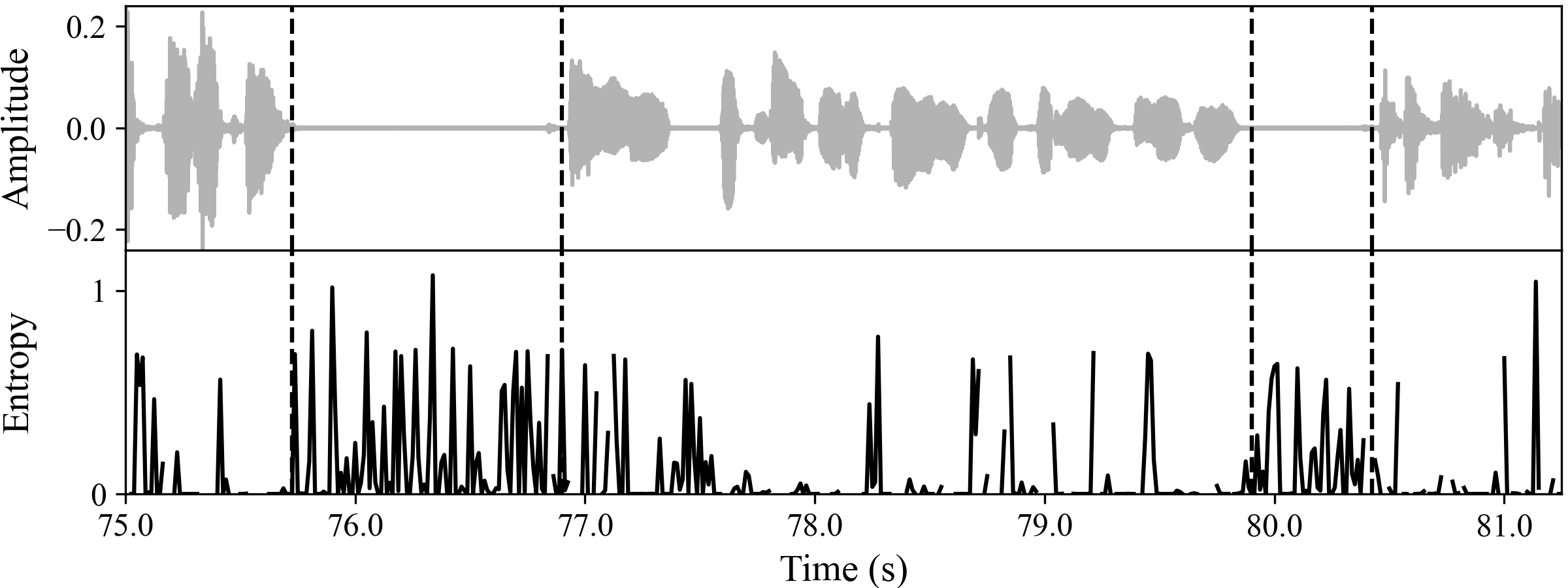}
    \vspace{-8mm}
    \caption{Frame level codebook entropy for AudioDec neural audio codec. High entropy is observed in silence regions.}
    \label{fig:codec-entropy}
\end{figure}

In this work, we propose using features from neural audio codecs (NACs) \cite{soundstream}. NACs encode speech into latent representations, followed by quantization, enabling low-bit-rate, discrete codes. We extract code vectors by indexing the codebook with the encoded codes and use these pre-trained code vectors as the input features for the LSTM endpointer. Figure \ref{fig:codec-entropy} shows the entropy\footnote{Frame-level entropy is computed from distance between encoder features and codebook vectors; code on GitHub repository.} of the first codebook for every frame with the AudioDec \cite{audiodec} codec. We can observe that the silence regions in speech tend to have high entropy in the codebook. This observation supports our hypothesis that pre-trained NACs possess features capable of differentiating between speech and silence regions, making them potentially helpful for endpointing.

Table \ref{tab:codec-params} shows different open-sourced, streaming NACs used. EnCodec \cite{encodec} and AudioDec \cite{audiodec} features are considered acoustic tokens while Mimi \cite{moshi} features are considered acoustic-semantic tokens \cite{ji2024wavchat}. We use pre-trained NACs at different frame rates, and with varying numbers of codebooks. Since EnCodec and AudioDec have high native frame rates, we also explore downsampled versions of their codec features. Causal average pooling is used for this downsampling process.

\subsection{Label delay}
\label{sec:label-delay}
To encourage the endpointer to delay its predictions, we introduce a delay to the labels during training. This induces an implicit latency learned from the modified training objective, but without requiring explicit latency during inference. Given the label sequence $\mathbf{Y} = (y_1, y_2, \dots, y_T)$ with $T$ frames, we generate the label sequence with delay $\tau$, $\mathbf{Y_\tau}$ as follows - 
    
\begin{equation}
\mathbf{Y_\tau} = 
\begin{cases}
    k & \text{if } t \leq \tau \\
    y_{t-\tau} & \text{if } t > \tau \text{ and } t \leq T \\
\end{cases}
\label{eq:label-delay}
\end{equation}

\begin{table}[t]
\caption{Configurations of different audio codecs. The parameter count only includes the encoder and quantizer parameters necessary for feature extraction. The frame rates obtained after downsampling codec features are shown in \textcolor{blue}{blue}.}
\vspace{-2mm}
\centering
\setlength{\extrarowheight}{1.5pt}
\scalebox{0.9}{

\begin{tabular}{l|c|c|c|c|c}
\hline
\makecell{Codec}                           & \makecell{params\\(M)}      & \makecell{frame\\ rate}  & \makecell{feature\\dim} & \makecell{single-stream \\endpointer\\params (M)} & \makecell{two-stream \\endpointer\\params (M)} \\ \hline
Spec & - & $25$ & $40$ & $2.1$ & $4.3$ \\ \hline 
EnCodec   & $7.43$ & $75$, \textcolor{blue}{$25$}&  $128$ & $2.4$ & $4.9$ \\
AudioDec  & $7.93$ & $80$, \textcolor{blue}{$20$}& $64$ & $2.2$ & $4.4$ \\
Mimi  & $38.34$ & $25$, $12.5$ & $512$ & $3.5$ & $6.9$ \\ \hline
\end{tabular}}
\vspace{-4mm}
\label{tab:codec-params}
\end{table}

To maintain consistent sequence lengths between $\mathbf{Y}$ and $\mathbf{Y_\tau}$, we prepend $\mathbf{Y}$ with $\tau$ instances of a special padding label $k$ and remove the final $\tau$ elements of $\mathbf{Y}$ as shown in Eq \ref{eq:label-delay}. During training, the loss is not computed for frames corresponding to the padding label, 
$k$. In our experiments, we trained models with delay values of 
$\tau$ up to $3$. With a frame rate of $25$ Hz, a delay of $\tau = 2$ corresponds to an implicit delay of $80$ ms  ($1000\times2/25$) in the training objective.

\section{Experimental setup}

\subsection{Dataset}
We train and evaluate our model using the offical train, validation and test splits of the SpokenWoZ corpus \cite{spokenwoz}. The dataset consists of many multi-turn spoken dialogues between two speakers (user and system), with minimal overlap in speaker turns. The corpus is characterized by pauses and filler words, thus creating a challenging and realistic endpointing scenario. It provides metadata with word-level timestamps, which we further pre-processed using Silero VAD \cite{silero} to remove leading and trailing silence from each turn. The original dataset contains 2-channel audio recordings sampled at 8kHz. For the two-stream endpointer (as depicted in Figure 1(b)), we utilize these two channels separately as distinct inputs for the user and system streams. Conversely, for the single-stream endpointer (Figure 1(a)), we generate a mono input by averaging the two channels.

\subsection{Training}
All models are trained using \textit{PyTorch} \cite{pytorch}, and the EnCodec and Mimi NACs are used from \textit{transformers}\footnote{\url{https://github.com/huggingface/transformers}} library \cite{transformers_lib}. Mel-spectrogram features are computed from the 8 kHz input audio using a $80$ ms window and $40$ ms shift. Since the Neural Audio Codecs (NACs) used in this study require 24 kHz audio, we resample the original 8 kHz audio to 24 kHz prior to NAC feature extraction. All NAC parameters are frozen; we train only the LSTM endpointer parameters. To account for varying input feature dimensions across different acoustic representations (see Table \ref{tab:codec-params}), LSTM parameters are optimized individually for each feature type. Specifically, the baseline (Mel-spectrogram), EnCodec, and AudioDec-based models use a 3-layer LSTM with 324-dimensional hidden features. The Mimi-based model employs a 2-layer LSTM with 512-dimensional hidden features. Additionally, the baseline model incorporates two non-linear projection layers preceding the LSTM, whereas the Mimi-based model uses one such layer.. All models were trained for $50$ epochs using frame-level cross-entropy loss and the Adam optimizer with a learning rate of $0.001$. The best model checkpoint is selected based on the average accuracy on the validation set over \texttt{$<$user$>$} and \texttt{$<$user-end$>$} predictions.


For the single-stream model (Figure \ref{fig:data-labels}(a)), we find it necessary to incorporate learnable embeddings for \texttt{$<$system$>$} tokens. As mentioned in Section \ref{sec:base_ep}, this addition is needed because the model must simultaneously discriminate between user and system speech segments while predicting their respective turn-end labels within a unified input stream. On the other hand, the two-stream endpointer (Figure \ref{fig:data-labels}(b)) inherently separates user and system processing through its dual-channel architecture, eliminating the need for explicit system activity tokens as the model can learn to distinguish speakers through separate input pathways.

As the dataset contains long dialogues, we create different training sequences of $40$ seconds in each iteration. We start the sequence at the start of a random \texttt{$<$user$>$} or \texttt{$<$system$>$} segment, and truncate it after $40$ seconds. During inference, we operate the endpointer on the entire test dialogue.

\subsection{Evaluation}
While general turn-taking prediction often focuses on accuracy improvements within fixed latency intervals \cite{ekstedt22_interspeech, morais2023modeling}, our work is interested in optimising across both error rate reduction and latency reduction, as minimising latency is crucial for user experience. Thus, we use three evaluation metrics commonly used in endpointing \cite{Shannon_Simko_Chang_Parada_2017}. We use \textit{ep50} and \textit{ep90} to measure latency, which corresponds to the time difference between the endpoint timestamp and the true turn-end timestamp. \textit{ep50} corresponds to the median latency, and \textit{ep90} represents the worst case - the tail latency at the $90$th percentile over all \texttt{$<$user$>$} turns. The cutoff error rate is captured by \textit{ep-cutoff} - the proportion of \texttt{$<$user$>$} turns where the endpointer is triggered before the true turn-end. Negative latency values are excluded from median and worst-case latency calculations, as early triggers are accounted for in the \textit{ep-cutoff} metric. We sweep the endpointer decision threshold probability between $0.7$ to $0.99$.

\subsection{Neural codec endpointer with speech LLM}
\label{sec:codec_llm}
Recent work has explored using endpointing or turn-taking predictors with speech LLMs to control duplex interaction \cite{zhang2025llm, chen2025minmo} and evaluate turn-taking \cite{talking_turns}. Building on this, we integrate our label-delay trained Mimi-based endpointer with Moshi \cite{moshi}, an open-sourced, Mimi NAC-based speech LLM. Notably, \cite{chen2025minmo, zhang2025llm} trained the endpointer jointly with the LLM on extensive conversational data; here, we integrate a pre-trained LLM.

We first establish a baseline by simulating interaction with Moshi and then describe the integration.\footnote{While this simulation uses Moshi, the control mechanisms using special tokens are applicable to other speech LLMs with similar architectures (e.g., \url{https://www.sesame.com/research/})}

\begin{figure}[t]
    \centering
    
    \includegraphics[width=1\linewidth]{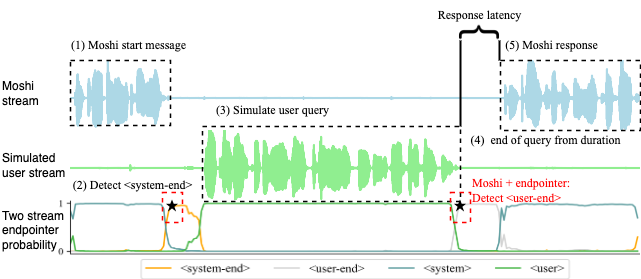}
    \vspace{-4mm}
    \caption{Our duplex simulation for Moshi speech LLM. Endpoint predicted from our real-time two-stream endpointer is denoted by $\star$.}
    \label{fig:full-duplex-sim}
    \vspace{-6mm}
\end{figure}

\noindent
\textbf{Simulating and Evaluating Moshi with Endpointer Control:}
This simulation measures Moshi's standalone response latency and cutoff rate (Figure \ref{fig:full-duplex-sim}).

\begin{enumerate}
    \item Moshi start message:
    Moshi initiates with a message (e.g., "How can I help you today?"). During this, the user stream is silent. Our two-stream endpointer concurrently monitors both streams and is predicting \texttt{$<$system$>$}.
    \item Detect \texttt{$<$system-end$>$}:
    We identify when Moshi system has completed it's initial message.
    \item Simulate user query:
    Upon detecting \texttt{$<$system-end$>$}, a 200-300 ms silence (simulating conversational pause) is introduced in the user stream, followed by a user query from the SpokenWoZ test set.
    \item For the baseline, detect end of query from the duration of simulated query segment
    \item Moshi response:
    Moshi's response latency is the duration from the query end to the start of Moshi's spoken response. A "cutoff" occurs if Moshi begins responding before the end of query.

\end{enumerate}

\begin{figure}[b]
    \centering
    \vspace{-6mm}
    \includegraphics[width=1\linewidth]{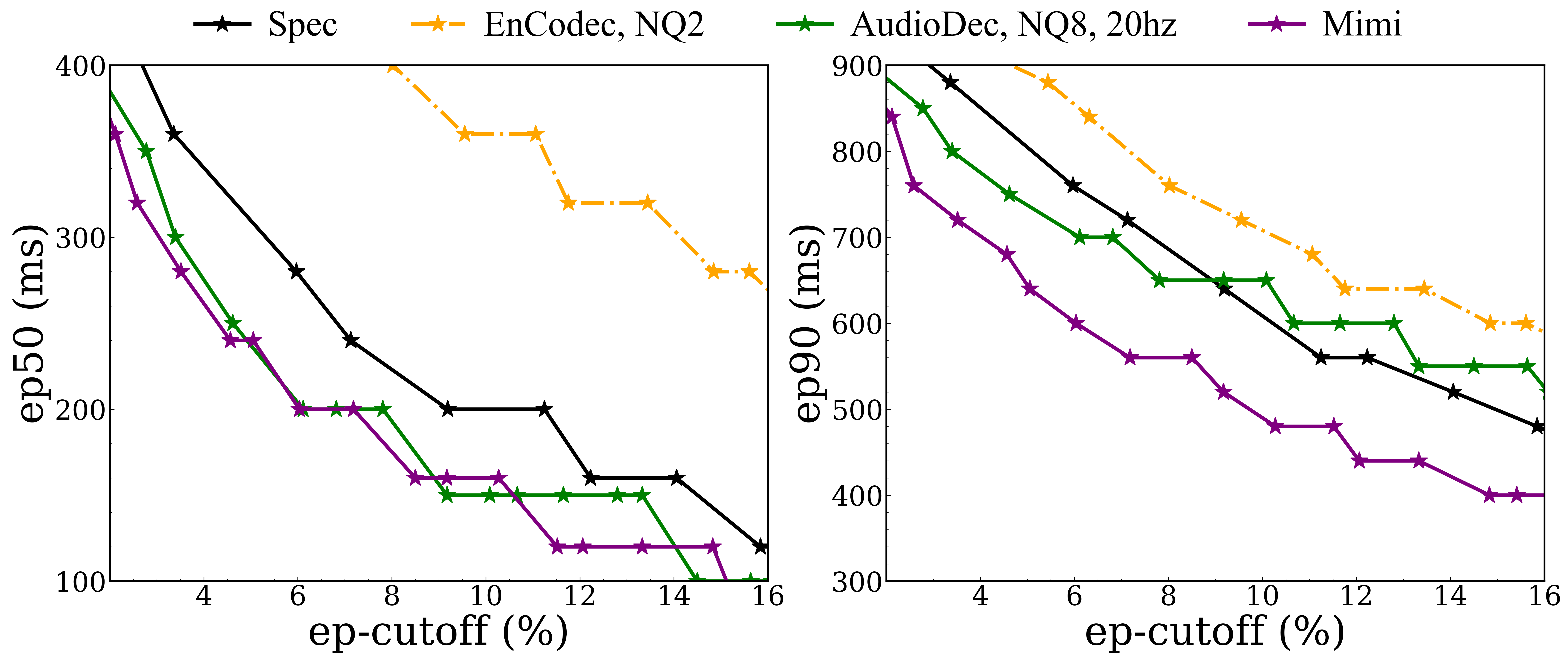}
    \vspace{-6mm}
    \caption{Single stream: Mel-Spectrogram-based endpointer (Spec) compared to neural audio codec-based endpointers. AudioDec is used at $20$ Hz, while the other endpointers are used at $25$ Hz.}
    \label{fig:codec-base}
\end{figure}

\noindent
\textbf{Endpointer-Controlled Moshi Interaction:}
This integration with Moshi serves to illustrate a key practical benefit of using NAC features for endpointing with compatible speech LLMs.To improve Moshi's responsiveness and reduce cutoffs, we integrate our Mimi-based endpointer. Notably, because both Moshi and our endpointer utilize Mimi NAC features, the underlying NAC encoder weights can be shared. This minimizes additional computational overhead, restricting new parameters primarily to the endpointer's LSTM layers. The endpointer operates in real-time, influencing Moshi's generation via its internal text stream (which guides speech output) using special tokens:
\begin{itemize}
    \item Preventing System Barge-in (Reducing Cutoff): While the user is speaking (during the simulated user query), if the endpointer detects user speech, we inject \texttt{<pad>} tokens into Moshi's internal text stream. This suppresses Moshi's speech generation, preventing it from interrupting the user.
    \item  Accelerating System Response (Reducing Latency): Immediately upon the endpointer's \texttt{$<$user-end$>$} trigger, we insert an \texttt{$<$unk$>$} token into Moshi's text stream. This prompts Moshi to begin generating its response from the subsequent step.
\end{itemize}

This evaluation uses 700 query-response pairs, selecting the first \texttt{<user>} segment exceeding one second from each dialogue in the SpokenWoZ test set. For further details on Moshi's inner monologue (text stream) and special token mechanisms, we refer the reader to \cite{moshi}. We rely on the default inference configurations in Moshi \footnote{\url{https://github.com/kyutai-labs/moshi}}. Our endpointer-Moshi integration code will be open-sourced.

\begin{figure}[t]
    \centering
    \includegraphics[width=1\linewidth]{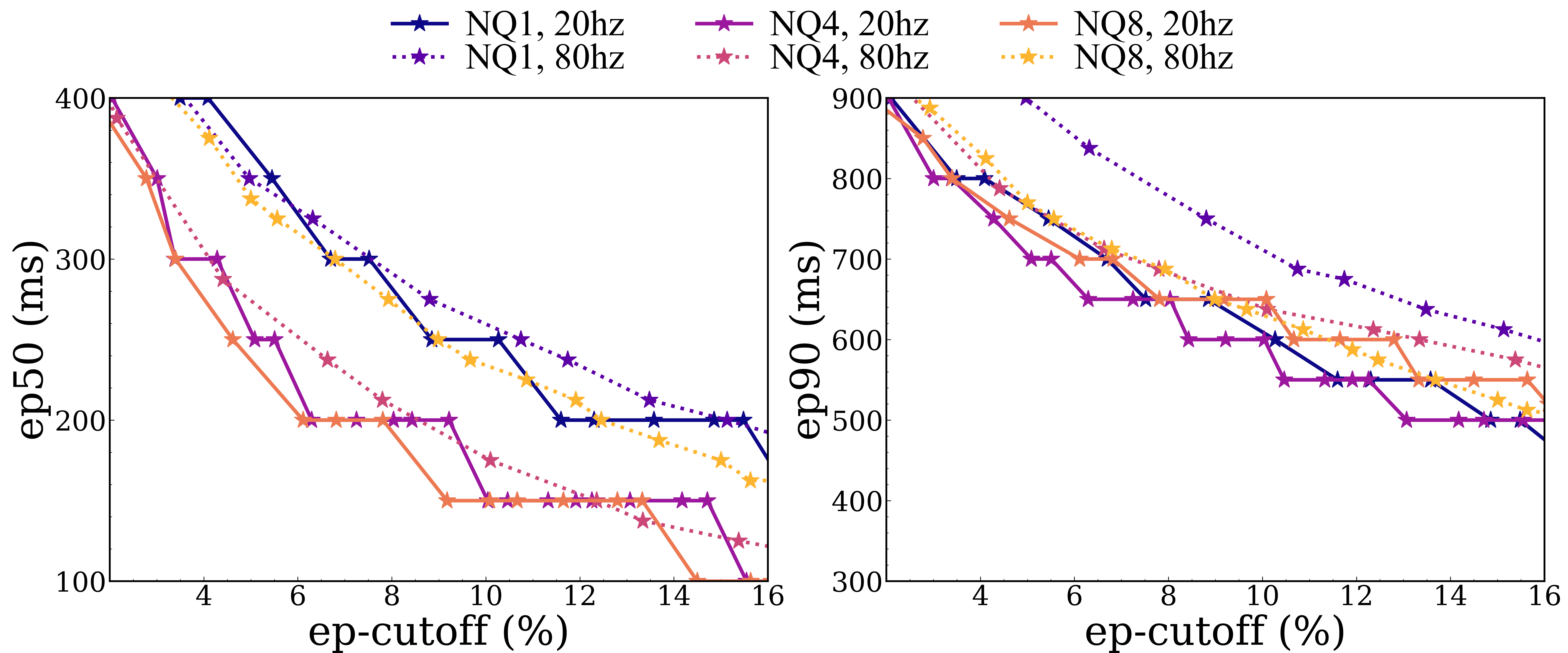}
    \vspace{-4mm}
    \caption{Single stream: Metrics on the AudioDec NAC-based endpointer, NQ corresponds to Number of Quantizer codebooks used.}
    \vspace{-6mm}
    \label{fig:audiodec}
\end{figure}



\section{Results and discussions}
\label{sec:results}
In this section, we will show the endpointer latency (\textit{ep50}, \textit{ep90}) and cutoff error rate (\textit{ep-cutoff}) for the baseline Mel-Spectrogram and proposed NAC based endpointers (Section \ref{sec:codecs}) from single-stream and two-stream models. Further, we will show the significance of the label delay training introduced in Section \ref{sec:label-delay}. Finally, we will provide an error analysis of the endpointer cutoff errors, followed by the evaluation of the codec-based endpointer with speech LLM (Section \ref{sec:codec_llm}).

\begin{figure}[b]
    \centering
    \vspace{-6mm}
    \includegraphics[width=1\linewidth]{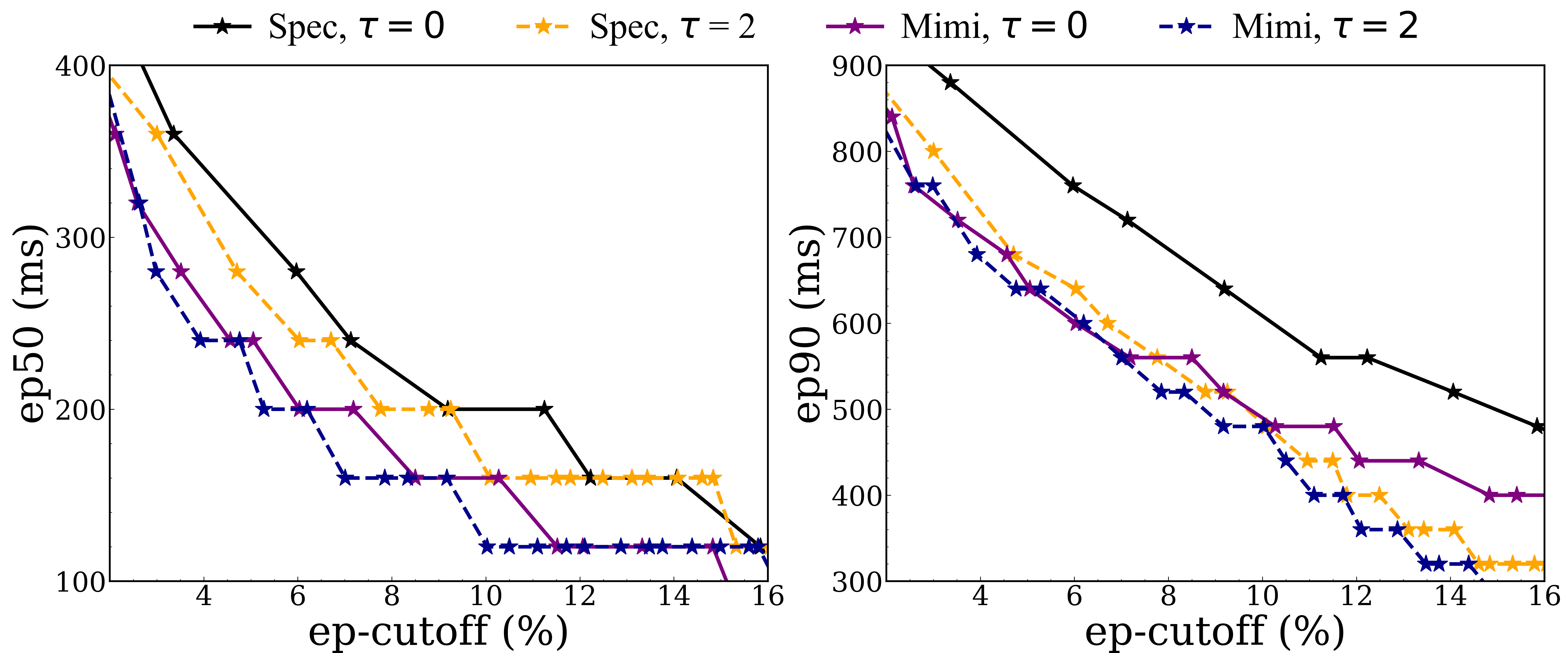}
    \vspace{-6mm}
    \caption{Single stream: Endpointer performance without label delay ($\tau = 0$), and with label delays ($\tau > 0$).}
    \label{fig:mel-delay}
\end{figure}

\subsection{Single-stream results}

\subsubsection{Neural audio codec performance}
Figure \ref{fig:codec-base} shows the metrics for endpointers using baseline Mel-Spectrogram versus proposed NACs as input. Mimi NAC achieves lower cutoff rates across both median and worst-case latencies, outperforming the baseline. This is likely due to the larger size of Mimi NAC (Table \ref{tab:codec-params}), and the semantic context distillation in the Mimi pre-training \cite{moshi}. AudioDec achieves similar error rates to Mimi when compared at median latency. However, it performs poorly against worst-case latency. As quantified in Table \ref{tab:overall-results} (comparing configurations with $\tau = 0$), at a 160 ms median latency,  the Mimi-based endpointer ($\tau = 0$) achieves an 8.50\% cutoff error, a 30\% relative reduction compared to the Mel-Spectrogram baseline's 12.23\% ($\tau = 0$).

Further investigation into NAC parameters (Figure \ref{fig:audiodec}, using AudioDec) indicates that lower input frame rates (20-25 Hz range) and an increased number of quantizer codebooks (e.g., NQ8 for AudioDec) generally lead to improved latency-accuracy trade-offs. This is likely due to the larger prediction context afforded by lower frame rates and the richer feature representation from more codebooks \cite{soundstream, audiodec}. These observations guided our choice of NAC configurations for subsequent experiments.

\subsubsection{Label delay}
Figure \ref{fig:mel-delay} shows the benefit of label delay training ($\tau > 0$) over standard training ($\tau = 0$). For both Mel-Spectrogram and Mimi features, label delay reduces cutoff errors across all operating points.  At a 160 ms median latency, (Table \ref{tab:overall-results}), the Mimi-based model achieves a cutoff rate of 7.01\%. This represents a significant 42.7\% relative error reduction compared to the baseline: Mel-Spectrogram without label delay (12.23\% for Spec, $\tau = 0$), showcasing the combined benefits of superior NAC features and label delay training. Furthermore, it also surpasses the Mel-Spectrogram endpointer with label delay (10.09\% for Spec, $\tau = 2$), confirming the advantages of Mimi features.

\begin{table}[t]
\caption{Single stream: Best endpointer results at different median latencies. Both Mel-Spectrogram and Mimi NAC features operate at $25$ Hz, Mimi uses all $8$ codebooks. Metrics are reported for two fixed median latency (ep50) operating points: values before the '/' correspond to an ep50 of 120 ms, and values after the '/' correspond to an ep50 of 160 ms.}
\vspace{-2mm}
\centering
\setlength{\extrarowheight}{3pt}
\resizebox{0.5\textwidth}{!}{
\begin{tabular}{c|c|c|c|c}
\hline
\makecell{Input} & \makecell{delay \\  ($\tau$)} & \makecell{worst-case latency \\ \textit{ep90} (ms)} & \makecell{cutoff error rate \\ \textit{ep-cutoff} (\%)} & \makecell{median latency \\ \textit{ep50} (ms)}  \\ \hline
Spec & $0$ & $480$ / $560$ & $15.84$ / $12.23$  & $120$ / $160$ \\                 
Mimi & $0$ & $480$ / $560$ & $11.52$ / $8.50$ & $120$ / $160$\\ \hline             
Spec & $2$ & $320$ / $480$ & $15.32$ / $10.09$  & $120$ / $160$ \\                       
Mimi & $2$ & $480$ / $560$ & $10.03$ / \textbf{7.01}  & $120$ / $160$ \\ \hline 
\end{tabular}}
\label{tab:overall-results}
\vspace{-6mm}
\end{table}

While the label-delayed Mimi endpointer shows higher worst-case latency than the similarly trained baseline at fixed median latencies, this trend reverses under fixed worst-case latency constraints. Specifically, at a 480 ms worst-case latency, the Mimi-based endpointer not only achieves a slightly lower error rate (10.03\% vs. 10.09\% for baseline) but also reduces median latency by 40 ms. This demonstrates that label delay training enhances Mimi's overall performance, offering a competitive advantage across various operating points.

\begin{figure}[b]
\vspace{-6mm}
    \centering
    \includegraphics[width=1\linewidth]{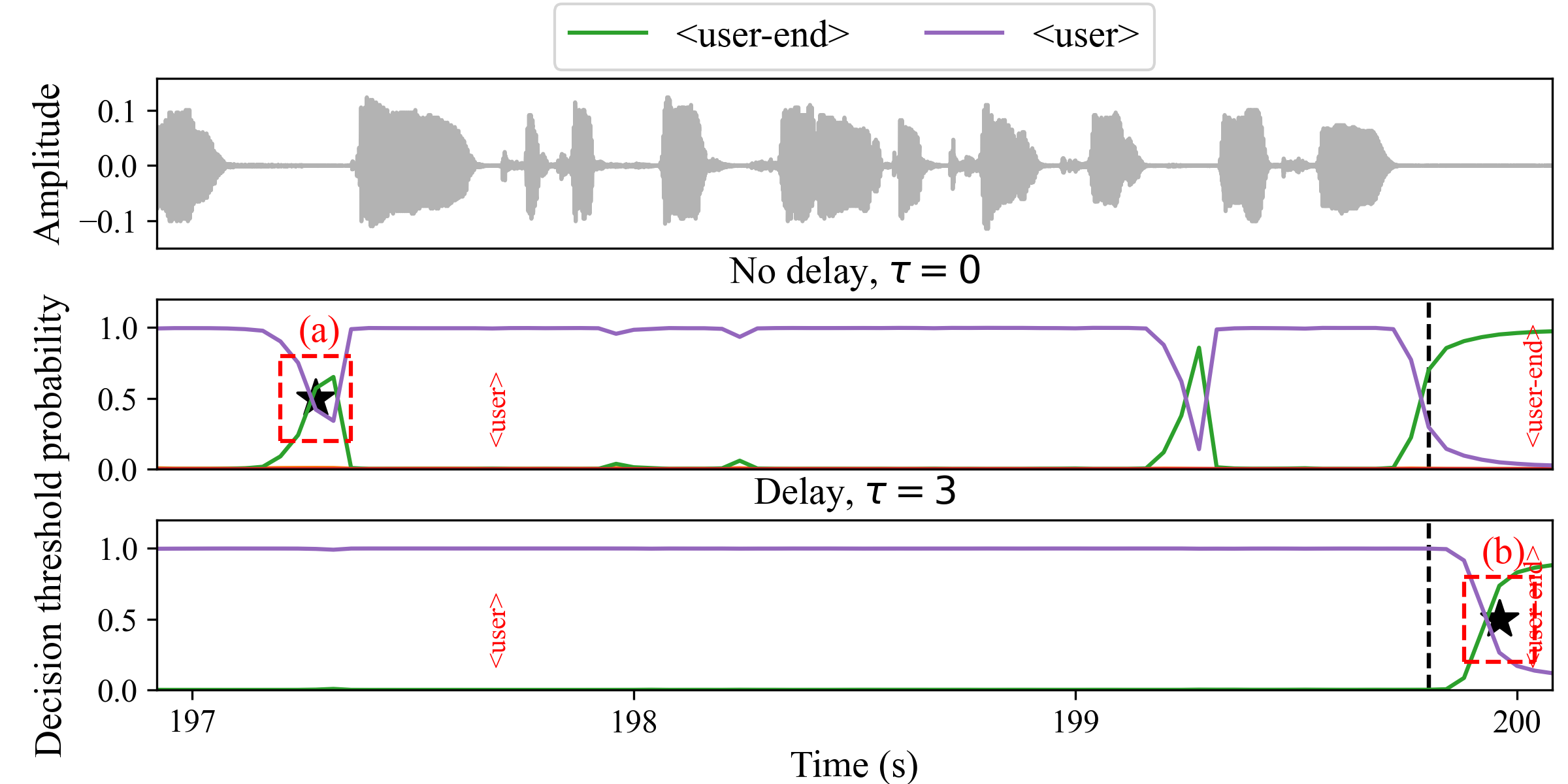}
    \vspace{-8mm}
    \caption{Endpoint triggered at $0.5$ threshold with and without delay training for single stream endpointer. $\star$ represents the estimated endpoint location, while the vertical dotted line indicates the ground truth endpoint.}
    \label{fig:post-plot}
\end{figure}

Figure \ref{fig:post-plot} demonstrates the advantage of label delay training in preventing premature endpointing errors. As shown in box (a), the baseline makes a premature error, while the label delay trained endpointer in box (b) correctly avoids this, highlighting the effectiveness of our approach.

\subsection{Two-stream results}

\subsubsection{Neural audio codec performance}
\begin{figure}[t]
    \centering
    \includegraphics[width=1\linewidth]{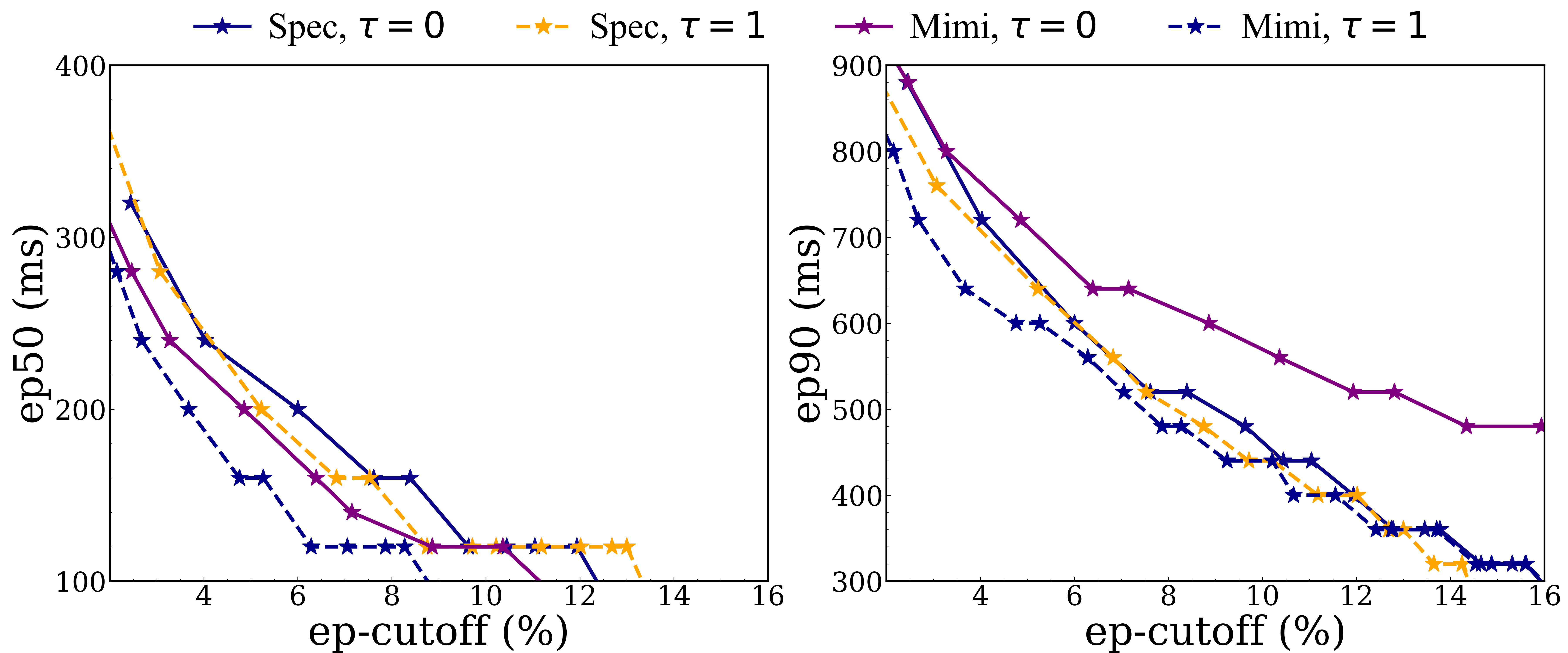}
    \caption{Two-stream: Endpointer performance without label delay ($\tau = 0$), and with label delays ($\tau > 0$).}
    \label{fig:ms-base-vs-codec}
    \vspace{-6mm}
\end{figure}

Figure \ref{fig:ms-base-vs-codec} presents results for the two-stream configurations, mirroring the trends observed in single-stream models: Mimi NAC features outperform Mel-Spectrograms, and label delay training further enhances performance for both feature types.

Our best two-stream configuration, the Mimi-based endpointer with label delay ($\tau = 1$), demonstrates strong performance as shown in Table \ref{tab:ms-overall-results}. At a 160 ms median latency, this model achieves a low cutoff error rate of 4.76\%. This is a 37.5\% relative error reduction compared to the two-stream Mel-Spectrogram baseline without label delay (7.61\% for Spec, $\tau = 0$). It also improves upon the Mel-Spectrogram baseline with label delay (6.82\% for Spec, $\tau = 1$), again underscoring the efficacy of combining Mimi NACs with label delay training. These results confirm that our proposed approaches generalize effectively to two-stream endpointing.

\begin{figure}[b]
    \centering
    \vspace{-6mm}
    \includegraphics[width=1\linewidth]{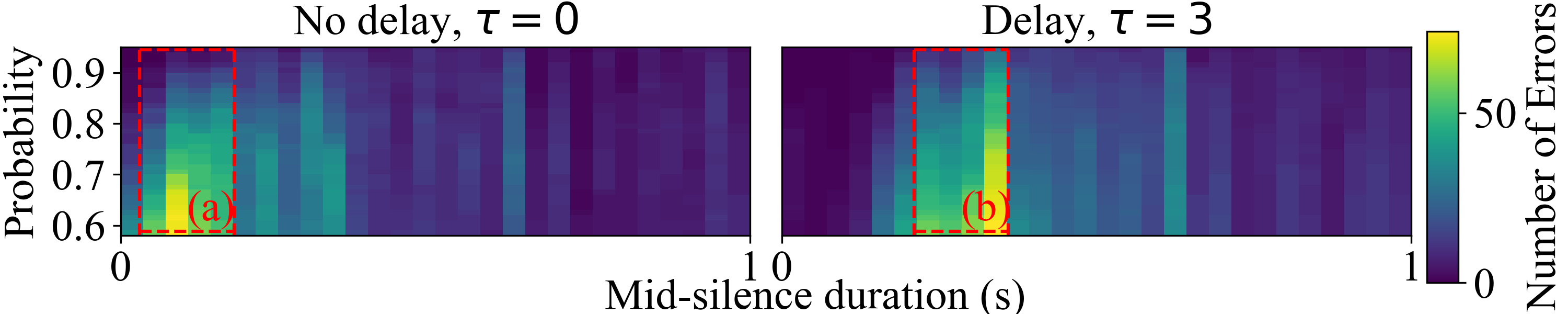}
    \vspace{-8 mm}
    \caption{Mid-silence durations corresponding to the cutoff errors encountered at different decision thresholds for endpointer with Mel-Spectrogram input feature.}
    \label{fig:error-locs}
\end{figure}

\subsubsection{Comparison with Single-Stream}
Comparing the best configurations at a 160 ms median latency, the two-stream Mimi model with label delay (4.76\% cutoff, 600 ms ep90, Table \ref{tab:ms-overall-results}) achieves a much lower error rate than its single-stream counterpart (7.01\% cutoff, 560 ms ep90, Table \ref{tab:overall-results}). This highlights the inherent advantage of processing user and system streams separately for multi-turn dialogues, although with a small increase in worst-case latency and model parameters (Table \ref{tab:codec-params}). We found that optimal label delays differed: $\tau = 2$ for the single-stream Mimi model and $\tau = 1$ for the two-stream version. This distinction likely reflects their differing architectural capabilities. The two-stream model, inherently more accurate at endpointing due to processing separate audio streams, needs only a smaller delay to reduce cutoff errors while maintaining responsiveness. Conversely, the single-stream model benefits from a relatively larger delay ($\tau = 2$) to better handle the ambiguities of mixed audio, leading to more reliable endpointing.

\subsection{Error analysis}

We analyzed premature cutoff locations, considering mid-silence durations determined by Silero VAD. As shown in Figure \ref{fig:error-locs}, without label delay, errors predominantly occurred during short mid-silences (box a). Label delay training reduced these short-pause errors but shifted some cutoffs to longer mid-silence regions (box b). This suggests label delay effectively prevents premature triggers on short pauses, though further work is needed for errors in longer silences.

\begin{table}[t]
\caption{Two-stream: Best endpointer results at different median latencies. Both Mel-Spectrogram and Mimi NAC features operate at $25$ Hz, Mimi uses all $8$ codebooks.}
\vspace{-2mm}
\centering
\setlength{\extrarowheight}{3pt}
\resizebox{0.5\textwidth}{!}{
\begin{tabular}{c|c|c|c|c}
\hline
\makecell{Input} & \makecell{delay \\  ($\tau$)} & \makecell{worst-case latency \\ \textit{ep90} (ms)} & \makecell{cutoff error rate \\ \textit{ep-cutoff} (\%)} & \makecell{median latency \\ \textit{ep50} (ms)}  \\ \hline
Spec & $0$ & 480 / $520$  &  9.64 / $7.61$ & 120 / $160$  \\                 
Mimi & $0$ & 600 / $640$ & 8.86 / 6.39  & 120 / $160$ \\ \hline             
Spec & $1$ & 480 / $560$  & 8.75 / $6.82$  & 120 / $160$  \\                       
Mimi & $1$ & 560 / $600$   & 6.28 / $\textbf{4.76}$ & 120 / 160  \\ \hline 
\end{tabular}}
\label{tab:ms-overall-results}
\vspace{-6mm}
\end{table}



\subsection{Using endpointer with Moshi speech LLM}
To demonstrate practical utility, we integrated our two-stream Mimi-based endpointer with the Moshi speech LLM. As Table \ref{tab:with-moshi} shows, this significantly improved performance; for example, at a 0.7 sampling temperature, median response latency dropped from 1640 ms to 412 ms, and cutoff error from 46.51\% to 21.15\%. This integration enables explicit response timing control without retraining the LLM and highlights the multi-task utility and computational efficiency of shared NAC features between endpointer and speech LLM.






\begin{table}[h]
\vspace{-2mm}
\caption{Endpointer with Mimi features ($12.5$ Hz, $8$ codebooks) with Moshi speech LLM at different sampling temperatures}
\vspace{-2mm}   
\centering
\addtolength{\tabcolsep}{-0.2em}

\setlength{\extrarowheight}{2pt}
\begin{tabular}{l|c|c|c|c}
\hline
Model config & \makecell{Sampling\\ temperature}  &    \makecell{Worst-case\\ latency (ms)}                                                    & \makecell{Cutoff \\rate (\%)} & \makecell{Median\\ latency (ms)}  \\ \hline

Moshi  & $0.7$ & $4575$ & $46.51$ &  $1690$    \\ 
Moshi  & $1.0$ & $4902$ & $56.86$ &  $1963$    \\ \hline
Moshi+endpointer & $0.7$ & $812$ & \textbf{21.15} & \textbf{412}   \\ 
Moshi+endpointer & $1.0$ & $\textbf{783}$ & 21.57 & $428$   \\  \hline
\end{tabular}
\label{tab:with-moshi}
\vspace{-4mm}
\end{table}

\section{Conclusions}
This work introduced streaming Neural Audio Codecs (NACs) for multi-turn dialogue endpointing, demonstrating that Mimi NAC-based models outperform traditional spectrum-based approaches. Our novel label delay training scheme significantly improved accuracy by better managing the latency-accuracy trade-off without added complexity. This combined approach yielded substantial relative cutoff error reductions (42.7\% single-stream, 37.5\% two-stream at 160 ms median latency) compared to a Mel-Spectrogram baseline. Error analysis confirmed label delay reduces premature triggers. Furthermore, integrating our endpointer with a speech LLM drastically cut LLM response latency and errors, highlighting its practical utility. Future work will extend this to general speech LLMs and incorporate multi-modal features for broader turn-taking tasks for open-domain conversations.

\bibliographystyle{IEEEtran}
\bibliography{template}

\end{document}